\documentclass[twocolumn,tighten]{aastex63}
\usepackage{amssymb, amsmath,framed}

\usepackage{bm}
\expandafter\ifx\csname package@font\endcsname\relax\else
 \expandafter\expandafter
 \expandafter\usepackage
 \expandafter\expandafter
 \expandafter{\csname package@font\endcsname}
\fi
\def\bq{\begin{equation}}
\def\eq{\end{equation}}
\def\bqy{\begin{eqnarray}}
\def\eqy{\end{eqnarray}}
\def\p{\partial}

\def\p{\partial}

\def\R{\mathbb{R}}

 \def\q#1#2{q^{\hspace{1pt}#1}_{\,,\hspace{1pt}#2}}
 \def\a#1#2{a^{\hspace{1pt}#1}_{\,,\hspace{1pt}#2}}
 \def\d#1#2{\delta^{\hspace{1pt}#1}_{\,,\hspace{1pt}#2}}

\begin{document}
\title{\large{Physical Constraints on Motility with Applications to Possible Life on Mars and Enceladus}}

\correspondingauthor{Manasvi Lingam}
\email{mlingam@fit.edu}

\author{Manasvi Lingam}
\affiliation{Department of Aerospace, Physics and Space Sciences, Florida Institute of Technology, Melbourne FL 32901, USA}
\affiliation{Institute for Theory and Computation, Harvard University, Cambridge MA 02138, USA}

\author{Abraham Loeb}
\affiliation{Institute for Theory and Computation, Harvard University, Cambridge MA 02138, USA}

\begin{abstract}
Motility is a ubiquitous feature of microbial life on Earth, and is widely regarded as a promising biosignature candidate. In the search for motile organisms, it is therefore valuable to have rough estimates for the number of such microbes that one may expect to find in a given area or volume. In this work, we explore this question by employing a simple theoretical model that takes into account the amount of free energy available in a given environment and the energetic cost of motility. We present heuristic upper bounds for the average biomass density and the number density of motile lifeforms for the Martian subsurface and the ocean of Enceladus by presuming that the motile microbes in question derive their energy from methanogenesis. We consequently demonstrate that the resultant densities of motile organisms might be potentially comparable to, or much lower than, the total microbial densities documented in various extreme environments on Earth.\\
\end{abstract}

\section{Introduction}\label{SecIntro}
Motility is a common characteristic of microorganisms on Earth. It is suspected that $\sim 20\%$ of all bacteria on Earth are motile \citep{Fen08}, but the exact fraction is uncertain by more than one order of magnitude \citep{AM07}. It has been hypothesized that motility is likely to constitute a generic feature of microbes in aquatic environments \citep[pg. 758]{NLD16}. There are sound reasons as to why the study of motility in generic astrobiological settings is desirable. For starters, the effective diffusion coefficient associated with motility is orders of magnitude higher than nonmotile organisms if the putative organisms have radii of $\gtrsim 1$ $\mu$m \citep[pg. 176]{Dus09}. On a related note, it is well known that motility, in tandem with chemotaxis, facilitates a substantive increase in nutrient uptake \citep[pg. 629]{Sto12}. 

At the same time, however, motility does incur significant energetic costs, which take up a considerable fraction of the total organismal metabolic budget \citep[pg. 678]{TS12}. Therefore, in theoretical studies that seek to assess habitability and the putative density of microbes from the standpoint of energetics \citep{TH07,SH07,HC20}, it is arguably necessary to incorporate these costs in a self-consistent fashion \citep{VanBo}. Last, from the perspective of \emph{in situ} life-detection missions, motility comprises a viable biosignature candidate \citep{NAD20,LL21,RSS21}. In fact, the spatial resolution that would be required to identify motile organisms is effectively lower compared with their nonmotile counterparts \citep[pg. 755]{NLD16}. 

It is thus valuable to place constraints on the maximum global biomass density and number density of motile microbes in a given habitat, as this can guide us in the design of life-detection experiments. Moreover, obtaining such estimates is helpful from the standpoint of empirically assessing the model because these constitute clear predictions that could be testable by future missions in principle. Hence, in this work, we select two representative habitable environments---the Martian subsurface \citep{MOM18} and the ocean of Enceladus \citep{HSHC}, both of which are regarded as being habitable and have been subjected to extensive analysis---and calculate heuristic upper bounds on the densities of motile organisms in these particular settings. 

The outline of the paper is as follows. In Sec. \ref{SecMot}, we provide a brief description of the model used to compute the power requirements for motility. Next, we apply a simple energetic model to estimate the maximal biomass and number density in Sec. \ref{SecMaxBio}. Finally, we summarize the salient findings and caveats in Sec. \ref{SecConc}.

\section{Model description for motile organisms}\label{SecMot}
Microbes on Earth have evolved a multitude of strategies for motility, each of which is attuned to the specifics of the habitats that they inhabit \citep{Berg}. Hence, we preface our analysis with the caveat that our results may have limited applicability in certain settings. Our results are, however, reasonably valid in microenvironments that are characterized by low Reynolds numbers; under such conditions, the prevalent mode of motility is taken to be run-and-tumble chemotaxis \citep[Section 4.2]{Kemp19}. To calculate the accompanying energetic cost, we will mirror the formalism presented in \citet{Mi02}; see also \citet{Berg} for additional quantitative details. In Appendix \ref{AppA}, we provide a detailed synopsis of the challenges associated with formulating quantitative models of motility and our rationale for working with run-and-tumble chemotaxis hereafter.

In the case of run-and-tumble chemotaxis, the total energetic cost of motility (units of W/cell), denoted by $P_m$, can be decomposed into two components as follows:
\begin{equation}\label{Ptot}
P_m = P_{rp} + P_{tp},    
\end{equation}
where the subscripts ``$r$'' and ``$t$'' signify the running (swimming) and tumbling (reorientation) phases, respectively. The expression for the swimming phase ($P_{rp}$) is given by the first term on the right-hand side of \citet[their Equation 6]{Mi02}, which is expressible as
\begin{equation}
    P_{rp} \approx \frac{k_B T D_s}{R^2},
\end{equation}
where $T$ is the temperature, $D_s$ denotes the molecular diffusivity of the solvent, and $R$ represents the radius of the cell. Over the habitable range of temperatures associated with life-as-we-know-it, both $T$ and $D_s$ vary by a factor relatively close to unity. Hence, we adopt the fiducial values of $D_s \approx 10^{-9}$ m$^2$ s$^{-1}$ and $T \approx 300$ K \citep{HHS00}. By substituting these estimates into the equation for $P_{rp}$, we arrive at
\begin{equation}\label{Prp}
    P_{rp} \approx 4.1 \times 10^{-18}\,\mathrm{W}\,\left(\frac{R}{1\,\mathrm{\mu m}}\right)^{-2}.
\end{equation}
By proceeding in a similar vein to calculate $P_{tp}$, we make use of the relations in \citet[pg. 730]{Mi02} and \citet[pg. 229]{Mi91}, which collectively yield
\begin{equation}\label{Ptp}
    P_{tp} \approx 3 \times 10^{-19}\,\mathrm{W}\,\left(\frac{R}{1\,\mathrm{\mu m}}\right)^{3}.
\end{equation}
Hence, by substituting (\ref{Ptp}) and (\ref{Prp}) in (\ref{Ptot}), we ascertain that $P_m$ becomes significant when $R \gg 1$ $\mu$m as well as when $R$ approaches the lower limit associated with microbes on Earth. The latter trend is consistent with the observation that $\gtrsim 10\%$ of the energy per unit time derived from metabolism is expended on motility by smaller bacteria \citep{Mi91}.

By calculating $d P_m/d R = 0$, the minimum is attained at $R_c \approx 1.6$ $\mu$m and the corresponding value of $P_m$ at $R = R_c$ is $P_c \approx 2.8 \times 10^{-18}$ W. As a point of comparison, the minimum power per cell required for maintenance, as predicted by both state-of-the-art models and empirical data, is $P_\mathrm{min} \approx 10^{-21}$ W \citep{BAA20}. There are two important points to bear in mind with regards to $R_c$ and $P_c$. First, neither of them are strictly constants, as they actually exhibit a weak dependence on $D_s$ and $T$. Second, for life-as-we-know-it, because neither of these two environmental parameters are subject to significant variation, it is conceivable that both $R_c$ and $P_c$ may evince a certain degree of generality.

\section{Maximum biomass for motile organisms}\label{SecMaxBio}
We can estimate the \emph{maximum} feasible biomass for chemoautotrophs ($M_\mathrm{max}$) by taking only the constraints from motility into account. The reason that we end up with an upper bound is because all the chemical energy is assumed to be utilized exclusively for the sake of motility in this simple model, and not toward the function of basal maintenance; the latter has a lower bound of $P_\mathrm{min}$. With this assumption, $M_\mathrm{max}$ is given by
\begin{equation}\label{Mmax}
    M_\mathrm{max} \approx \frac{\mathcal{C}_\mathrm{cell}\, |\Delta G|\, \Phi_\mathrm{max}\, A_\mathrm{bio}}{P_\mathrm{bio}},
\end{equation}
where $P_\mathrm{bio}$ represents the temporally averaged power required for a motile microbe, $\mathcal{C}_\mathrm{cell}$ is the mass of the organism, $\Delta G$ is the net Gibbs free energy associated with the exergonic chemical reaction in the given environmental conditions, $\Phi_\mathrm{max}$ denotes the maximum flux of the appropriate reactant(s), and $A_\mathrm{bio}$ is the cross-sectional area of the putative biosphere \citep{SBH13,SKC19,LiLo20}. 

The fraction of time spent by microbes in a state of motion is denoted by $\epsilon$. The temporally averaged power expended by microbes on this activity is therefore roughly given by $P_\mathrm{bio} \approx \epsilon P_m$. We adopt a characteristic value of $\epsilon \sim 0.1$, as many bacteria are known to evince $\epsilon \lesssim 0.2$ \citep[pg. 12]{MK06}. In select environments, $\epsilon \ll 1$ could certainly be manifested---especially in energy-limited habitats that stymie the prospects for motility \citep{LRL15}---but we are interested in exploring the characteristics of moderately active and motile microbes; for these reasons, we shall hereafter employ $\epsilon \sim 0.1$. Next, we are free to adopt $\mathcal{C}_\mathrm{cell} \approx 4\pi \rho_w R^3/3$, where $\rho_w$ is the density of water; this relation follows from the datum that the density of cells is close to that of water \citep{MP16}.

\begin{figure}
\includegraphics[width=8.0cm]{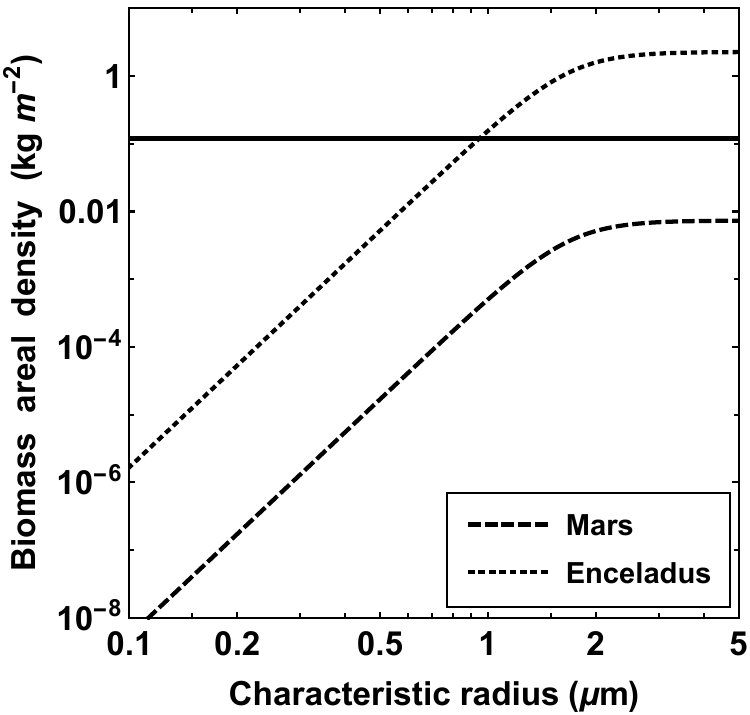} \\
\caption{The upper bound on the total biomass of motile organisms per unit area (biomass areal density) in SI units of kg m$^{-2}$ as a function of the microbe radius (in $\mu$m) for putative methanogens in the Martian subsurface and the ocean of Enceladus. These estimates were calculated for the fiducial choice of $\epsilon \sim 0.1$ in (\ref{sigmax}), which is the fraction of time spent in motion. The horizontal black line is an estimate of the biomass areal density of Earth's deep biosphere.}
\label{FigMotCost}
\end{figure}

By plugging the above relations in (\ref{Mmax}), we arrive at
\begin{equation}
   M_\mathrm{max} \propto \left[4.1 \left(\frac{R}{1\,\mathrm{\mu m}}\right)^{-5} + 0.3 \right]^{-1},
\end{equation}
where it is now necessary to interpret $R$ as the characteristic radius of the ensemble of microbes present in a particular habitat. It is apparent that $M_\mathrm{max}$ decreases monotonically with $R$, implying that most of the biomass is concentrated in the form of microbes with sizes of at least a few microns. This behavior stands in contradistinction to $P_m$, because the latter is a nonmonotonic function of $R$, as demonstrated earlier. In contrast, let us suppose that we wish to calculate the number of microbes; this requires us to divide $M_\mathrm{max}$ with $\mathcal{C}_\mathrm{cell}$. After doing so, the result is inversely proportional to $P_m$. Hence, the upper bound on the number of motile cells attains a maximum when $P_m$ is minimized: that is, when $R = R_c$ and $P_m = P_c$. In other words, the number of motile cells is maximized when the putative microbes exhibit a radius of $\sim 1.6$ $\mu$m, as explained in Sec. \ref{SecMot}.

The other noteworthy aspect is that $M_\mathrm{max} \propto A_\mathrm{bio}$, as seen from (\ref{Mmax}), indicating that larger habitats/worlds are likely to host more biomass. Hence, it is more instructive to analyze the biomass areal density (i.e., vertically integrated biomass per unit area), defined as $\sigma_\mathrm{max} = M_\mathrm{max}/A_\mathrm{bio}$. To calculate $\sigma_\mathrm{max}$, it is imperative to consider a specific pathway and location. For the sake of comparison, we consider putative methanogen-like microbes. Our reasons are twofold. First, the reactants necessary for methanogens (viz., CO$_2$ and H$_2$) are widely available in many astrobiological environments, as the ensuing examples illustrate. Second, methanogens are capable of survival and growth in diverse conditions and have therefore served as model organisms in many experimental and theoretical analyses that were designed to emulate or simulate extraterrestrial settings \citep{MS05,TS15,TP18,HC20,LL21}.

The first setting that we consider is the Martian subsurface, because this was investigated in \citet{SKC19}. Another point of significance is that liquid water mixtures have been confirmed in this environment \citep{Wray}. On the basis of the simulations and data presented in \citet{SKC19}, we normalize $\Phi_\mathrm{max}$ by $\Phi_M \approx 1.3 \times 10^{12}$ m$^{-2}$ s$^{-1}$ and $\Delta G$ by $\Delta G_M \approx - 24$ kJ mol$^{-1}$, with the subscript ``$M$'' embodying the the Martian subsurface. After invoking (\ref{Mmax}), we end up with
\begin{eqnarray}\label{sigmax}
&& \sigma_\mathrm{max} \approx 2.2 \times 10^{-3}\,\mathrm{kg\,m^{-2}}\,\left(\frac{\Phi_\mathrm{max}}{\Phi_M}\right) \left(\frac{|\Delta G|}{|\Delta G_M|}\right)
\left(\frac{\epsilon}{0.1}\right)^{-1} \nonumber \\
&& \hspace{0.5in} \times \left[4.1 \left(\frac{R}{1\,\mathrm{\mu m}}\right)^{-5} + 0.3 \right]^{-1}.   
\end{eqnarray}
In contrast, we note that Earth's deep biosphere has been estimated to comprise a total biomass of roughly $20$ Pg C \citep[pg. 712]{MLD18}, which would translate to a total biomass of $\sim 6 \times 10^{13}$ kg, after using a dry-to-total biomass conversion factor of $3$ \citep{BD84}. Hence, this translates to $\sigma_\oplus \approx 1.2 \times 10^{-1}$ kg m$^{-2}$ for Earth's deep biosphere after employing a cross-sectional area of $4\pi R_\oplus^2$. It is not surprising that (\ref{sigmax}) is a few orders of magnitude smaller than $\sigma_\oplus$, because the former is derived under the assumption of a relatively high $\epsilon$ as elucidated earlier. On the other hand, if one considers $P_\mathrm{bio} \sim P_\mathrm{min}$ (which effectively amounts to $\epsilon \ll 1$), corresponding to primarily sessile microbes such as those ostensibly existent in the deep biosphere, it is straightforward for $\sigma_\mathrm{max}$ to be comparable to $\sigma_\oplus$.

In Figure \ref{FigMotCost}, we have plotted $\sigma_\mathrm{max}$ for the Martian subsurface and for the ocean of Enceladus as a function of $R$. The reason for choosing the latter environment is that molecular hydrogen was confirmed in the plumes of Enceladus by the \emph{Cassini} mission \citep{WGP17}. We adopt $|\Delta G| \approx 80$ kJ mol$^{-1}$ and $\Phi_\mathrm{max} \approx 1.2 \times 10^{12}$ m$^{-2}$ s$^{-1}$ on the basis of the combined empirical constraints and geochemical modeling presented in \citet{WGP17}. Note that $\Phi_\mathrm{max}$ is higher than $\Phi_M$ by nearly two orders of magnitude because Enceladus is anticipated to have active hydrothermal processes. 

As a consequence, it is evident from Figure \ref{FigMotCost} that $\sigma_\mathrm{max} > \sigma_\oplus$ for $R > 1$ $\mu$m, although it is a few times smaller than Earth's surface biomass density of $\sim 3.2$ kg m$^{-2}$, the latter of which is derived after invoking the biomass estimates from \citet{BPM18}. It is apparent from the figure that $\sigma_\mathrm{max}$ increases with the size until it eventually saturates, in accordance with (\ref{sigmax}). The lower bound of $R$ in all the figures is specified to be $\sim 0.1$ $\mu$m due to biophysical constraints \citep{Mor67,MWJ15}, although it may be a few times higher when additional constraints on gradient sensing are taken into account \citep{Dus09,LM21}.

At this stage, an important point is worth reiterating. We have only examined the energetic costs directly stemming from the act of locomotion. In other words, we have not studied the auxiliary costs involved with motility like the manufacture of flagella and cilia. Empirical evidence suggests that the majority of the energetic costs for reasonably active microbes are attributable to locomotion (i.e., they are enshrined in $P_m$), implying that our analysis might represent a reasonable approximation for this class of organisms (see \citealt{TS12,KVB17}).

\begin{figure}
\includegraphics[width=8.0cm]{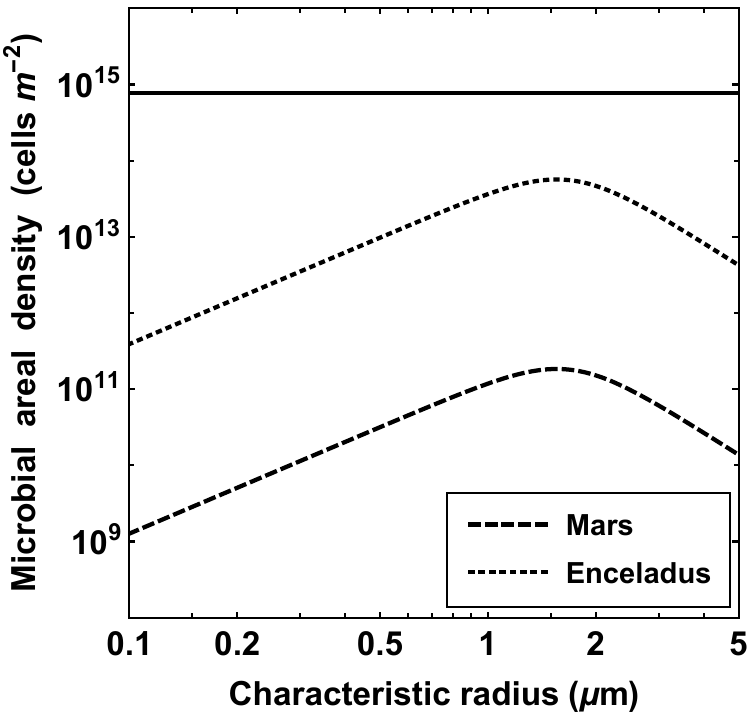} \\
\caption{The upper bound on the column number density (i.e., microbes per unit area) of motile microbes in SI units of cells m$^{-2}$ as a function of the microbe radius (in $\mu$m) for putative methanogens in the Martian subsurface and the ocean of Enceladus. These estimates were calculated for the fiducial choice of $\epsilon \sim 0.1$ in (\ref{etamax}), which is the fraction of time spent in motion. The horizontal black line is an estimate of the column density of Earth's deep biosphere.}
\label{FigAreaDen}
\end{figure}

Instead of plotting the biomass areal density, it is possible to calculate the column number density of cells $\eta_\mathrm{max}$ (units of cells $m^{-2}$) instead. This quantity is determined by dividing $\sigma_\mathrm{max}$, which is given by (\ref{sigmax}), by $\mathcal{C}_\mathrm{cell}$. As a consequence, we end up with
\begin{eqnarray}\label{etamax}
&& \eta_\mathrm{max} \approx 5.2 \times 10^{11}\,\mathrm{cells\,\,m^{-2}}\,\left(\frac{\Phi_\mathrm{max}}{\Phi_M}\right) \left(\frac{|\Delta G|}{|\Delta G_M|}\right) \nonumber \\
&& \hspace{0.4in} \times \left(\frac{\epsilon}{0.1}\right)^{-1} \left[4.1 \left(\frac{R}{1\,\mathrm{\mu m}}\right)^{-2} + 0.3 \left(\frac{R}{1\,\mathrm{\mu m}}\right)^{3} \right]^{-1}.   
\end{eqnarray}
For comparison, when it comes to Earth's deep biosphere, it contains approximately $4 \times 10^{29}$ cells \citep[pg. 707]{MLD18} over an area of $4\pi R_\oplus^2$. Hence, the column number density of microbes in Earth's deep biosphere is $\eta_\oplus \approx 7.8 \times 10^{14}$ cells m$^{-2}$. For the fiducial values chosen above, it is apparent that $\eta_\mathrm{max} \ll \eta_\oplus$, but this condition does not hold true when $\epsilon$ drops several orders of magnitude below unity. 

Figure \ref{FigAreaDen} illustrates $\eta_\mathrm{max}$ as a function of $R$ for the Martian subsurface and the ocean of Enceladus, along the lines of Figure \ref{FigMotCost}. There are a couple of interesting points that emerge from this figure. First, for the appropriate choice of $\epsilon$, we notice that $\eta_\oplus$ is comfortably higher than $\eta_\mathrm{max}$ for all values of $R$. As explained previously, this trend is not surprising given that most microbes in Earth's deep biosphere are sessile and characterized by $P_\mathrm{bio} \sim P_\mathrm{min}$. Second, $\eta_\mathrm{max}$ is a nonmonotonic function of $R$ and attains a maximum value at $R = R_c$, the latter of which was derived earlier. The reason stems from the fact that $\eta_\mathrm{max} \propto 1/P_m$ in our model.

Last, if the motile microbes are assumed to be uniformly mixed throughout the typical column depth $H$ of the solvent, the number density of microbes $\mathcal{N}_\mathrm{max}$ can be estimated as follows:
\begin{equation}\label{rhomax}
  \mathcal{N}_\mathrm{max} \approx \frac{\eta_\mathrm{max}}{H}.
\end{equation}
It is, however, not easy to gauge the magnitude of $H$ for the Martian subsurface because what matters is not the total depth of the subsurface habitable region, but the fraction of that depth composed of the solvent in which the putative microbes dwell, which is hard to ascertain. If we consider the ocean of Enceladus, it may be a reasonable approximation to suppose that, over the span of geological timescales, microbes inhabit the entirety of its ocean, thereby amounting to $H \approx 40$ km \citep{BRT16,HM19}. 

Hence, in the case of Enceladus, by adopting the above value of $H$ and making use of (\ref{etamax}) and (\ref{rhomax}), it is possible to calculate $\mathcal{N}_\mathrm{max}$. For the basis of comparison, it is instructive to invoke the total microbial number densities associated with some of the harsher environments on Earth. Subglacial lakes (e.g., Lake Gr\'imsv{\"o}tn in Greenland and Lake Vostok in Antarctica) exhibit densities of $\gtrsim 10^8-10^{10}$ cells m$^{-3}$ \citep[Table 1]{PM16}, the deep-ocean sediments of the North Pacific Gyre are characterized by $\sim 10^9$ cells m$^{-3}$ \citep{OSKE,RKA12}, and groundwater extracted from granitic rocks $\sim 3$ km below the surface evinces densities of $\gtrsim 10^{10}$ cells m$^{-3}$ \citep[Table 1]{LWR06}. 

\begin{figure}
\includegraphics[width=8.0cm]{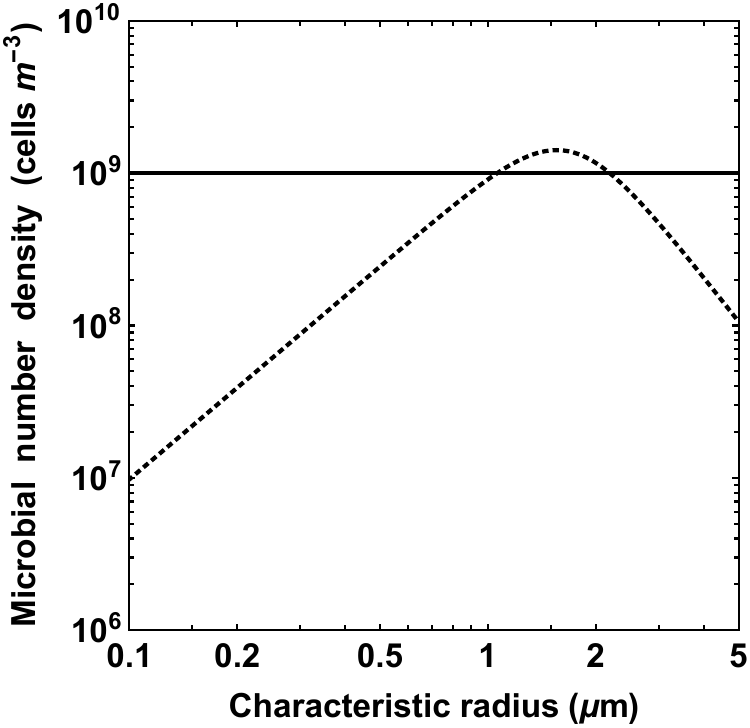} \\
\caption{The upper bound on the number density (i.e., microbes per unit volume) of motile microbes in SI units of cells m$^{-3}$ as a function of the microbe radius (in $\mu$m) for putative methanogens in the ocean of Enceladus. These estimates were derived for the fiducial choice of $\epsilon \sim 0.1$, which is the fraction of time spent in motion. The horizontal black line embodies the characteristic \emph{total} (i.e., motile and sessile) microbial number densities documented for extreme environments on Earth.}
\label{FigNumDen}
\end{figure}

On the basis of the above paragraph, a cell density of $\rho_\oplus \approx 10^9$ cells m$^{-3}$ is a good measure of the \emph{total} (i.e., motile and sessile) microbial number density in extreme habitats on Earth. In Figure \ref{FigNumDen}, we have plotted $\mathcal{N}_\mathrm{max}$ as a function of $R$ and contrasted it against $\rho_\oplus$. We caution that $\mathcal{N}_\mathrm{max}$ is exclusively for motile organisms, whereas $\rho_\oplus$ is composed of both motile and immobile microbes. There are a couple of aspects that stand out. First, on account of the fact that $\mathcal{N}_\mathrm{max} \propto 1/P_m$, it follows that the number density attains a maximum at $R = R_c$. Second, and more importantly, we discover that $\mathcal{N}_\mathrm{max}$ is close to $\rho_\oplus$ under near-optimal circumstances, but is otherwise roughly an order of magnitude (or more) smaller than the latter. Hence, this would suggest that the average density of motile microbes in the ocean of Enceladus might end up being very low. 

In closing, we note that it is feasible to calculate the mass density of microbes by dividing $\sigma_\mathrm{max}$ with $H$ and plugging in the numbers of Enceladus. We have, however, opted not to do so because most studies tend to focus on determining or estimating $\mathcal{N}_\mathrm{max}$, both in Earth-based and extraterrestrial habitats.

\section{Conclusion}\label{SecConc}
It is well-known that motility represents a genuine biosignature candidate \citep{NLD16}, and that it accords microbes a number of adaptive benefits. Hence, motivated by these considerations, we explored the prospects for motility in astrobiological environments. 

We began by showing that run-and-tumble chemotaxis, the \emph{de facto} mode of motility under many circumstances, is rendered optimal in terms of power consumption when the microbes have radii of $\sim 1.6$ $\mu$m. Next, by making use of a simple bioenergetic model wherein the vast majority of the available free energy is directed toward enabling motility, we assessed the maximum amount of biomass that could exist in the form of motile organisms. In particular, we calculated the global integrated biomass density per unit area $\sigma_\mathrm{max}$, the column number density $\eta_\mathrm{max}$ (number of microbes per unit area) and the number density $\mathcal{N}_\mathrm{max}$ (number of microbes per unit volume) of microbes, with the last result following from the additional postulate of uniform mixing. 

We applied our results to the Martian subsurface and to the ocean of Enceladus. We focused on methanogens as these organisms have been well studied on Earth and they require access to molecular hydrogen, which is believed to be prevalent in both locales. Among other results, we showed that the estimates for $\eta_\mathrm{max}$ for both Mars and Enceladus are likely to be lower than the corresponding value for Earth's deep biosphere. We also demonstrated that $\mathcal{N}_\mathrm{max}$ in the ocean of Enceladus is potentially smaller than the total number density of microbes across myriad harsh environments on Earth. 

As with any model, there are some potential caveats in need of highlighting. First, our analysis was conducted for a particular variant of motility (run-and-tumble chemotaxis). Adopting a different mode would give rise to changes in the scalings, although the magnitudes of the relevant variables might remain similar \citep{Mi02}. Second, our model set aside the metabolic requirements for basal maintenance and the synthesis of swimming appendages (e.g., flagella and cilia), which is why all of our predictions constitute upper bounds; the actual values may consequently end up being much smaller. 

Third, as our model was predicated on tackling only energetic constraints, it ignored the possibility that nutrient limitations could act to suppress the biomass further. More specifically, theoretical modeling by \citet{LiMa18,MaLi19} indicates that phosphorus limitation might engender additional bottlenecks insofar as Enceladus is concerned. Fourth, our estimates represent global averages, and therefore do not account for spatial heterogeneity, for example, hotspots wherein free energy sources and/or nutrients are concentrated and lead to the clustering of microbes. In order to model the actual density of motile organisms at such locations, it would be necessary to modify $\Phi_\mathrm{max}$ and $|\Delta G|$ accordingly. However, this remains a challenging enterprise in the current epoch, given our limited empirical data regarding the (sub)surface environments of habitable worlds in the Solar system.

In spite of these ostensible drawbacks, our model nevertheless has two major advantages, which were adumbrated in Sec. \ref{SecIntro}. The first is that it enables us to at least achieve a rough understanding of the global cell densities of mobile microorganisms to anticipate in future astrobiological missions and to plan for such eventualities. Second, our results are transparent due to the relative simplicity of the model and can therefore be readily assessed and falsified by forthcoming life-detection experiments. It is thus our hope and expectation that this work will pave the way for quantifying the prospects for motility in extraterrestrial milieu.

\acknowledgments
We are grateful to Chris McKay and an anonymous reviewer for their meticulous and insightful reviews of the paper, which were helpful for improving the manuscript. This research was supported in part by the Breakthrough Prize Foundation, Harvard University's Faculty of Arts and Sciences, and the Institute for Theory and Computation (ITC) at Harvard University.

\appendix

\section{Challenges and rationale for modeling run-and-tumble chemotaxis} \label{AppA}
Many studies that have attempted to gauge the energetic costs associated with basic physiological functions of microbes tend to use run-and-tumble locomotion on account of its applicability to environments with low Reynolds numbers \citep{Berg,KVB17,Kemp19}. Although run-and-tumble chemotaxis is often perceived as being apropos for nutrient-rich habitats, there are two caveats that should be recognized. 
\begin{enumerate}
    \item Observational data concerning the availability and abundance of bioessential nutrients in most astrobiological environments is very scarce. For example, while the Cassini mission collected evidence favoring the existence of nitrogenous compounds during its passage through the plume of Enceladus \citep{PKN18}, the abundance of phosphorus-based species remains unknown due to the absence of empirical data. Hence, it is not inconceivable that some of these settings have more nutrients than expected and are primarily energy-limited.
    \item Run-and-tumble chemotaxis is anticipated to be functional even in low-nutrient environments, such as those inhabited by halophilic archaea on Earth \citep{TBD20}.
\end{enumerate}
Looking beyond the issue of nutrient abundance, there are multiple chemotactic strategies that are accessible. There are, however, other uncertainties that arise in this context, which are explicated below.
\begin{enumerate}
    \item There are substantive differences in how motility is effectuated in bacteria and archaea, but our understanding of swimming and other related properties of the latter is still much less understood than the former \citep{HW12}.
    \item If we specialize to bacteria, there are a diverse array of chemotactic strategies that are feasible such as run-and-tumble, run-and-arc, run-and-reverse, run-tumble-flick, and run-reverse-flick. 
    \item If we further consider only a single parameter, even in this restricted case, we are confronted with much heterogeneity. For instance, increased phosphorylation of the CheY protein inhibits tumbling in gram-positive \emph{Bacillus subtilis}, whereas the converse is valid for gram-negative \emph{Escherichia coli} \citep{GO95}.
    \item In connection with the above point, the swimming speed of a given organism can vary by a factor of order unity owing to its propensity to modulate its locomotion when travelling forward or backward, as seen from the example of the soil bacterium \emph{Pseudomonas putida} \citep{TTZ13}.
\end{enumerate}

In light of the preceding considerations, it may be argued that a first attempt at estimating the biomass density of motile microbes should attempt to construct a model endowed with sufficient simplicity and generality. A heuristic model can subsequently pave the way for more complex analyses of this pertinent subject, as contended in Sec. \ref{SecConc}. It is apparent that the basic version of run-and-tumble chemotaxis---which applies to a spherical swimmer---has the requisite property of simplicity \citep{Kemp19}. Now, when it comes to addressing the question of generality, it is necessary to compare run-and-tumble locomotion with other strategies; we will mostly draw upon the results derived in \citet{Mi02}, except when stated otherwise. 
\begin{enumerate}
    \item For organisms with radii of $\gtrsim 1$ $\mu$m, the length and aspect ratio of the flagellum do not significantly influence the power requirements for run-and-tumble locomotion; in other words, the power expended varies merely by a factor of order unity.  
    \item For organisms characterized by $R \gtrsim 0.8$ $\mu$m, the power requirements for run-and-tumble locomotion are rendered almost independent of the length scales associated with the chemical gradients pertaining to chemotaxis.
    \item The power requirements for run-and-tumble and run-and-arc locomotion are within an order of magnitude of each other, provided that the radius of the microbe is $\lesssim 10$ $\mu$m. 
    \item In the range of $R \gtrsim 0.5$ $\mu$m, it is plausible that the run-and-tumble and run-and-reverse modes of locomotion have comparable power requirements.
    \item The mean squared displacement for run-and-tumble, run-and-reverse, and run-reverse-flick is within an order of magnitude even over a timescale that is $\sim 100$ times the mean run time \citep{TSZ13}.
    \item For many bacteria (e.g., \emph{Escherichia coli}), the predicted chemotactic drift speeds are similar for run-and-tumble, run-tumble-flick, and run-reverse-flick modes of locomotion \citep{TSZ13}.
\end{enumerate}
As indicated by these points, run-and-tumble locomotion might possess a certain degree of generality, implying that the results derived in the paper may also be imbued with some validity. Furthermore, the range of radii described above overlap, for the most part, with the radii at which the biomass variables in Figures \ref{FigMotCost}--\ref{FigNumDen} reach their maximal values. We reiterate, however, that our analysis is not meant to be definitive and could thus serve as a stepping stone for more sophisticated treatments of this subject in the future.

%\bibliographystyle{aasjournal}
%\bibliography{Motility}

\end{document}